# Development of Compact Multivariable Probe for Two-Phase Detection in PbLi-Argon Columns

Abhishek Saraswat
*Institute for Plasma Research*
Gandhinagar, India
*Indian Institute of Technology Madras*
Chennai, India
asaraswat@ipr.res.in

Ashokkumar Prajapati
*Institute for Plasma Research*
Gandhinagar, India
ashokp@ipr.res.in

Rajendraprasad Bhattacharyay
*Institute for Plasma Research*
Gandhinagar, India
rbhattac@ipr.res.in

Paritosh Chaudhuri
*Institute for Plasma Research*
Gandhinagar, India
*Homi Bhabha National Institute*
Mumbai, India
paritosh@ipr.res.in

Sateesh Gedupudi
*Indian Institute of Technology Madras*
Chennai, India
sateeshg@iitm.ac.in

*Abstract*—Liquid-gas two-phase flow is a common occurrence in various industrial applications. However, for nuclear fusion applications with a Li based liquid-breeder, existence of two-phase flow may lead to critical issues including decreased Tritium Breeding Ratio (TBR), generation of hot-spots and improper nuclear shielding. Additionally, a very large density ratio of Liquid-Metal (LM) to gas mandates relevant experimental database towards development and validation of software tools. Out of the candidate liquid-breeders, lead-lithium eutectic (Pb-16Li) has gained immense focus for its various advantages and is utilized in several breeding-blanket concepts. In view of the above-mentioned requirements, it is imperative that a LM–gas two-phase detection tool be developed for high-temperature liquid PbLi environment. For two-phase detection in LMs, electrical-conductivity based probes are most suitable in terms of ruggedness, ease of fabrication and simplicity of operation. However, corrosive nature coupled with high operational temperature for PbLi facilities puts severe demands on electrical-insulation, a foremost requirement for electrical-conductivity based detection schemes. In this study, a multivariable probe based on electrical-conductivity and temperature measurement schemes is developed using an electrical-insulation coating of high-purity alumina ($Al_2O_3$). Developed probe is validated in high-temperature PbLi-Ar two-phase vertical column with bulk PbLi temperature upto 400°C and time-averaged void-fraction upto 0.95 covering flow regimes from dispersed bubbly flow upto in-box Loss of Coolant Accident (LOCA) characterized by a very large gas flow inside bulk PbLi. Developed probe provides high reliability with excellent temporal-resolution towards individual bubble detection using electrical-conductivity based principle alongwith simultaneous two-phase temperature trends. Present paper provides details about sensor probe fabrication and calibration, LM-gas two-phase test-facility, time-averaged void-fraction estimations using threshold method, bubble-frequency and bubble residence-time estimations alongwith critical observations from the preliminary experimental investigations.

## I. INTRODUCTION

Nuclear fusion breeder-blanket concepts employ attractive solid and liquid breeder materials in the form of lithium/lithium-containing compounds like Li, Pb-16Li, $Li_2O$, $LiAlO_2$, $Li_4SiO_4$, $Li_2SiO_3$, $Li_2TiO_3$ and $Li_2ZrO_3$ [1-2]. Out of these candidate materials, Pb-16Li (hereafter referred to as PbLi) has gained immense focus for its various advantages including a high TBR without an additional neutron-multiplier, circulation ability facilitating tritium extraction outside blanket, immunity towards radiation damage and thermal stresses, high thermal-conductivity and reduced chemical-activity compared to pure Li [2]. Success of a breeder concept is primarily governed by TBR and heat-extraction performance, which can be well achieved using PbLi in a self-cooled concept. However, interaction of Li with fusion neutrons to breed tritium leads to generation of helium gas as a by-product, which has a low-solubility in PbLi and could precipitate in the form of bubbles affecting system design and safety [3-5]. Entrapped gases within breeder/coolant LM circuit may cause local hot-spots, improper shielding and a reduction in TBR. For PbLi flow-rates upto 1000 kg/s, gas-phase generation is significant and independent of pressure [3]. Various concepts like HCLL (Helium-cooled Lithium Lead), WCLL (Water-cooled Lithium Lead), DCLL (Dual Coolant Lithium Lead) and LLCB (Lead-lithium Ceramic Breeder), therefore, would invariably be prone to such a phenomenon. In some of the concepts (HCLL and LLCB), an in-box Test Blanket Module (TBM) LOCA will lead to ingress of high-pressure helium gas inside PbLi circuit resulting in a liquid-gas two-phase flow with high density ratio between two-phases. Preliminary experimental studies at Institute for Plasma Research (IPR) with water as a test-fluid corroborated presence of trapped gas-pockets at 90° bends and entrained gas-bubbles in re-circulation zones inside TBM-like complex geometry. Similarly, two-phase flow and trapped gas-pockets are also expected in lab-scale R&D facilities due to LM charging in presence of inert cover gas. To model such occurrences of relevance towards design and operational safety of ancillary breeder/coolant circuits for future fusion reactors, extensive experimental database needs to be generated, mandating development of proper diagnostic tools compatible with high-temperature and corrosive PbLi environment. Numerous experimental studies for room temperature and low-melting LMs have been conducted worldwide utilizing commercial and specialized techniques [6-12] like PIV, laser, γ-ray, X-ray, neutron emission etc. Although such techniques benefit from non-intrusive nature of detection [13-14], the opaque nature, extreme operating environments, installation constraints, requirements of localized detections and high-attenuation characteristics of LM render most of the techniques challenging towards practical implementation. As a preliminary attempt to study two-phase flow regimes, electrical-impedance based techniques offer a better route considering ease of installation, feasibility of adaptation and good response owing to large difference in electrical-conductivities of LM and gas. A detailed two-phase study on PbBi eutectic utilizing electrical-conductivity probe has been conducted upto a temperature of 200°C [15]. However, adaptation to PbLi puts severe demands on electrical-insulation compatibility towards corrosive media and operational temperature upto 400°C. Considering such limitations and unavailability of experimental data, studies have been recently initiated with numerical tools to predict different two-phase flow regimes in PbLi environment [16]. As stated in recent literature [16] and to the best knowledge of authors, no experimental data exists on two-phase flow detection for PbLi. This study primarily aims to bridge the existing gap with development and preliminary validation of a measurement tool to study two-phase flow regimes in PbLi environment. In this work, gas-phase detection in bubbly flow regime with estimations of time-averaged void-fraction ($\alpha$) has been performed for bulk temperature upto 400°C. Validation of measurement technique is further extended towards accidental scenarios like in-box LOCA simulated with controlled Ar gas injection at high flow-rates varying from 1.5 slpm to 3.5 slpm in vertical PbLi column.

## II. Materials and Methods

### A. Probe Design

The multivariable probe is fabricated from a specifically designed and calibrated duplex configuration K-type thermocouple with outer sheath of 1.5 mm diameter and sensing junctions located 1.5 mm from sheath tip. Ungrounded hot-junction configuration allows utilization of single probe for simultaneous electrical-conductivity based measurements and bulk-temperature measurements. Dip-coating technique is applied, on the probe-sheath except the tip, to coat a thin layer of $AlPO_4$-bonded calcined $Al_2O_3$ with 99.8% purity (characterized as pure $\alpha$-$Al_2O_3$ phase) available as refractory coating suspension. Probe-sheath is then inserted into a recrystallized alumina tube (200 mm length, 2 mm bore and 1 mm wall thickness) intended to provide protection and ruggedness against impacts from Ar bubbles rising in high-density LM column. End of the tube (near sensor-tip) is closed by filling same coating suspension, such that ~1.5 mm of thermocouple sheath is uncoated (exposed). After a 24 h air-set period, heat-cure performed in a muffle furnace, at each coating stage, includes following steps:

- Gradual dehydration: Ramp-up from room temperature (RT) to 93°C over 2 h followed by heat-cure at 93°C for 2 h.
- Bonding to substrate: Ramp-up from 93°C to 427°C over 12 h followed by heat-cure at 427°C for 2 h.
- Natural cool-down from 427°C to RT.

After first heat-cure, annular region between sensor sheath and tube wall is filled with coat suspension using a hypodermic needle and second heat-cure is performed. Closure of tube rear-end with application of coating over outer surface on 100 mm of the tube-length followed by third heat-cure completes probe fabrication. Relevant section of final fabricated probe is shown in Fig. 1. Working principle for the probe is similar to that described under [17] except that the second electrode is replaced by another duplex K-type thermocouple, which also measures bulk PbLi temperature. Therefore, fabricated probe provides thermal trends for two-phase flow in addition to discrete voltage dips corresponding to bubble interactions with exposed metallic-tip. Average diameter of as-fabricated probe is ~ 4.6 mm corresponding to coating thickness ~300 μm. Insulation leakage current ($I_L$) was < 1 nA at RT with a test voltage of 275 VDC (IR tester Make: Fluke; Model: 1550C). In addition to facilitating installation of probe in test tank, utilized compression fittings provide gas-tight seal to maintain inert environment and minimize probe movement during bubble impaction.

### B. Test-Facility: Design and Operation

Fig. 2(a) represents schematic of designed test-facility (without PbLi tank) highlighting PbLi wetted components. Assembled test set-up is shown in Fig. 2(b). Thermal gradients within bulk PbLi are minimized using diameter and height of the tank as 52.48 mm and 100 mm respectively. PbLi wetted components of facility are fabricated of SS-304 while probe-sheath is made of SS-316. Melting of PbLi chunks and thereafter temperature regulation is achieved within ± 5°C using feedback temperature control based resistive surface heating. A constant inert cover environment is maintained with positive Ar pressure and repeated Ar flushing. A separate bare-sheathed thermocouple dipped inside bulk PbLi acts as reference electrode for electrical-conductivity based circuit while also providing bulk PbLi temperature before generation of two-phase flow. Bubble interaction probability with probe sensing-tip is increased by means of Ar-gas injection using specially designed 8-legged spider configuration sparger with three bores of 2 mm in each leg. During continuous Ar injection, momentary closure of vent valve with simultaneous rise in system pressure ensures gas flow in facility. Subsequent to achieving steady Ar flow, sparger is inserted in PbLi till tank bottom, from where the Ar gas injected nearly at RT and after traversing through complete axial length of sparger, emerges at the bottom of molten PbLi column and rises upwards generating a two-phase flow-regime. After stabilization, hybrid probe is inserted and contact with PbLi is ensured using real-time voltage measurements and simultaneous rise of both the temperature signals of the multivariable probe. At low injection flow-rates, variation in Ar influx is achieved using a manual needle-valve downstream of Ar gas cylinder. At higher flow-rates, a Digital Mass Flow Controller (DMFC) is employed to realize $\alpha$ similar to in-box LOCA accidental conditions. Distance between the probe-tip and top layer of sparger legs is ~55 mm.

## III. Results and Discussions

### A. Calibration and Response-Time Estimations

Probe calibration is performed using manual dip and retrieval cycles in static PbLi at bulk temperature of 290°C. High electrical conductivity of PbLi [18-19] generated pulsed voltage output nearly equal to the excitation voltage of 5 VDC. Sampling of thermocouple raw-signals (mV) was performed synchronous to conductivity-circuit voltage signal while offline temperature conversion is performed using constant Seebeck coefficient of 40μV/°C as applicable in the operating range of interest. Error minimization for cold-junction compensation is achieved using ambient control at Data-Acquisition (DAQ). Utilized PXIe-6363 module, with an absolute accuracy of ⩽33 μV, provides an absolute error less than ± 1°C. Further processing for temperature trends involves smoothening using a moving-average function for last 250 samples. A representative calibration data logged at 0.2 kHz sampling frequency is shown in Fig. 3. From t = 0 to 5.63s, probe is outside PbLi and shows a zero-voltage output due to open circuit condition. As soon as the probe is dipped in PbLi (t = 5.63s), pulsed output of ~5V is obtained and simultaneous sharp positive slope is observed for both temperature trends leading towards bulk PbLi temperature. It should be noted that rise in temperature between t = 0 to 5.63s is the effect of probe travel towards PbLi before immersion. At t = 48.45s, probe is pulled out of PbLi leading to a zero voltage corresponding to break in continuity with simultaneous decreasing trend in both temperature values. The time-duration from t = 48.45s to 77.68s simulates a plug flow trend (not to scale). Similar pattern is repeated for probe insertion at t = 77.68s and retrieval at t = 92.62s. Similarity of temperature trends corroborates accuracy of measurements from both junctions of multivariable probe. From t = 106.89s to t = 115.66s, output response towards repeated dip/retrieve cycles with approximately 1s dip-time, highlights excellent response of the electrical-conductivity based measurements but sluggish response on part of temperature signals. However, detailed observation suggests magnitude of temperature during this period, i.e. ~283°C, remains in-between the temperature magnitudes when probe-tip is in liquid-phase (~289°C) and when probe-tip is in gas-phase (~276°C). Detailed analysis of interpolated voltage signal for one of the immersion/retrieval cycles, shown in Fig. 4(a) and (b), establishes $T_{90}$ response-time of ~4.9 ms for rising voltage (immersion in PbLi) and ~4.5 ms for dropping voltage (retrieval from PbLi) i.e. simulating bubble impaction using electrical-conductivity principle. Finite rise-time and drop-time arise for voltage signal due to entrained gas-volume near the sensing-tip during sudden immersion and adhered thin-film of PbLi during retrieval, respectively [20]. Although, previous work has demonstrated clear dependance of calibration and response-time on sampling-frequency [17], a relatively lower sampling at 0.2 kHz suffices for conclusive and qualitative real-time detection in this case considering high electrical-conductivity of PbLi compared to water.

*B. Criterion for Conclusive Detection*

Assuming the case of an irregular shaped single bubble with impaction-chord length = $l + d$, as shown in Fig. 5 (where $l$ = length of conducting-tip of probe) rising vertically in PbLi column with velocity $v$. As discussed in [17], voltage dip initiates after complete encapsulation of conducting-tip (AB) by the bubble and voltage rise will commence as soon as length $d$ has been traversed across point B. In limiting case, where the time taken to traverse $d$ exactly equals $T_{90}$ response-time for voltage dip case, conclusive bubble detection based on $T_{90}$ response is given by:

$$d = vT_{90} \qquad (1)$$

Therefore, with a sensing-tip of 1.5 mm, bubbles with impaction-chord lengths of 1.6 mm and 2 mm could be detected conclusively using $T_{90}$ threshold, if traversing probe-tip at velocities upto 2.22 cm/s and 11.11 cm/s, respectively.

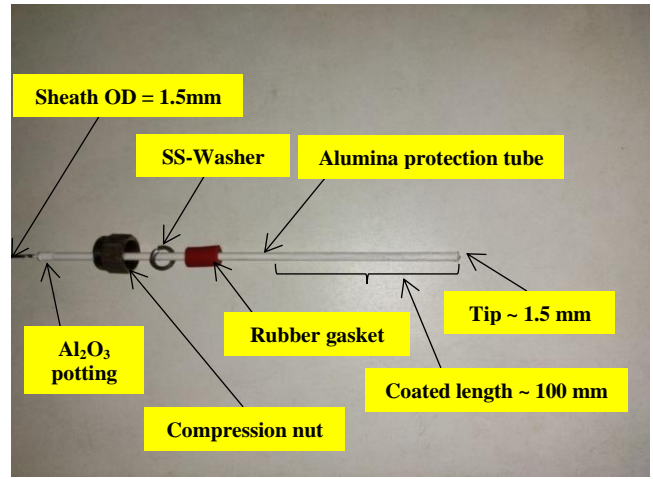

Fig. 1. As-fabricated probe for PbLi-Ar two-phase detection

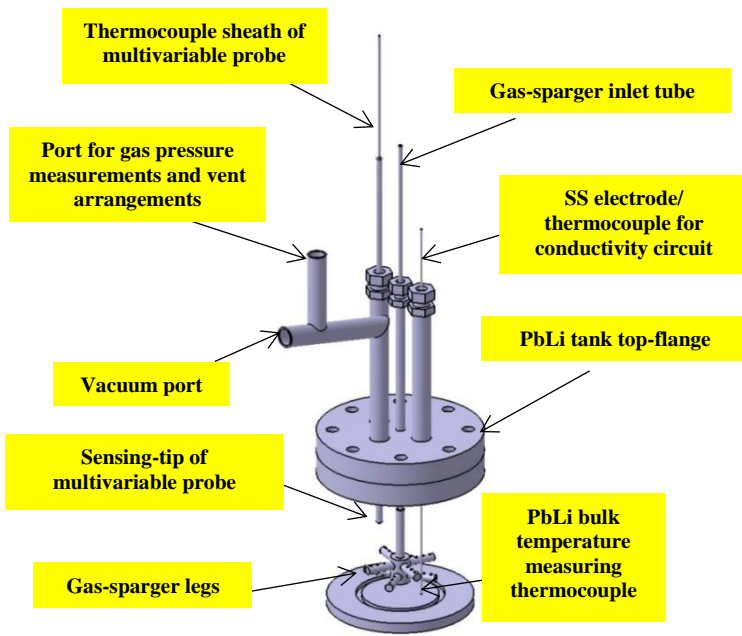

Fig. 2(a). Schematic representation of test-facility for Ar (gas) bubbling in high-temperature PbLi (liquid)

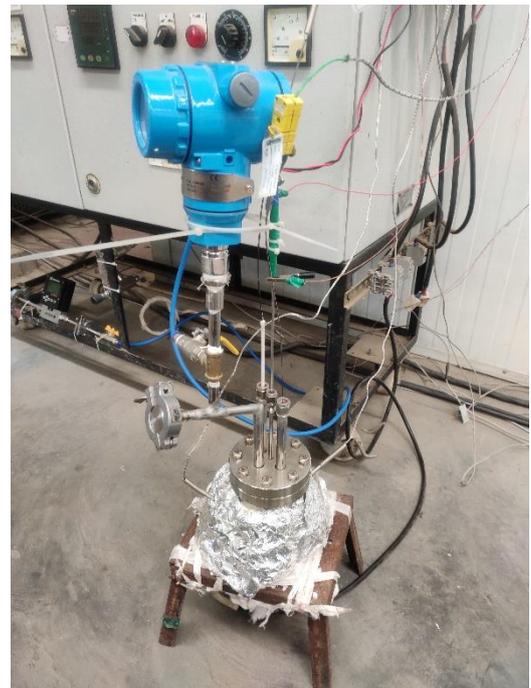

Fig. 2(b). Assembled two-phase flow set-up with installed probe (same orientation as in Fig. 2(a))

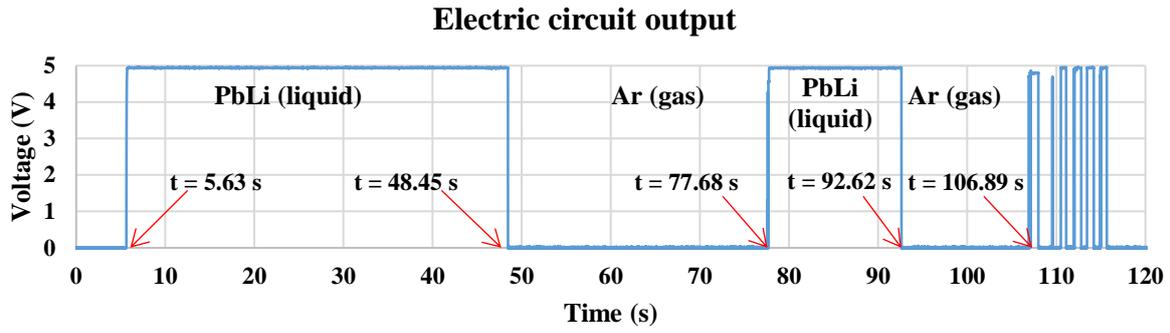
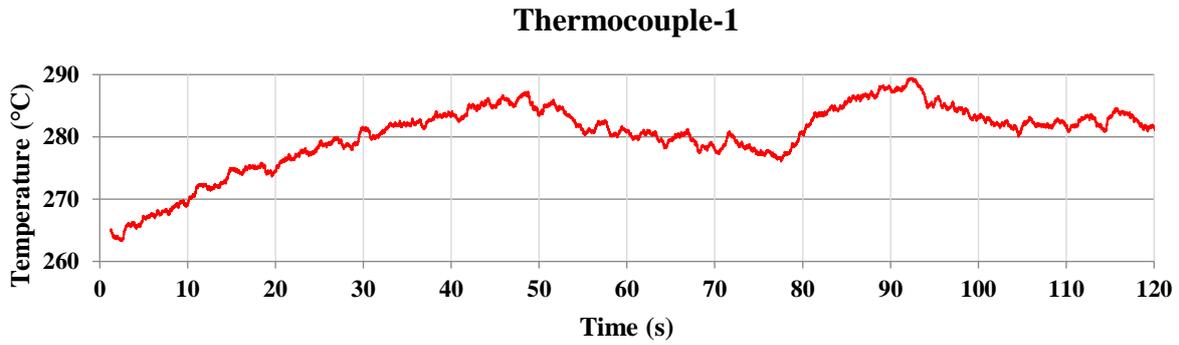
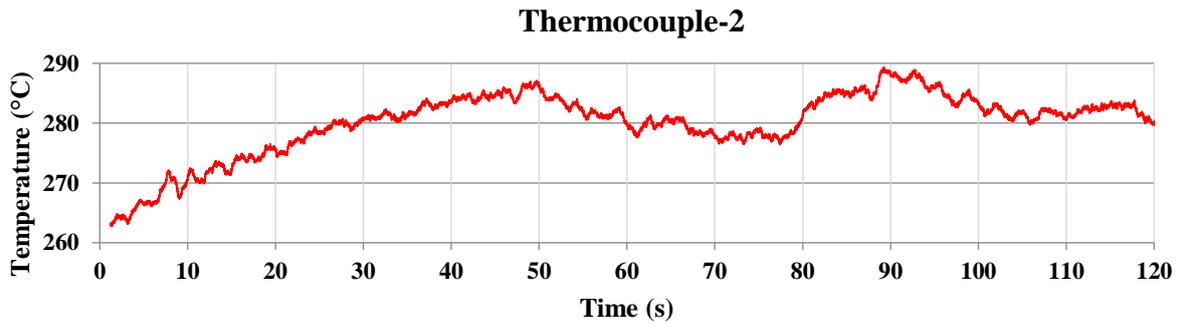

Fig. 3. Calibration test highlighting the effect of PbLi contact on the electrical-conductivity based signals and temperature trends

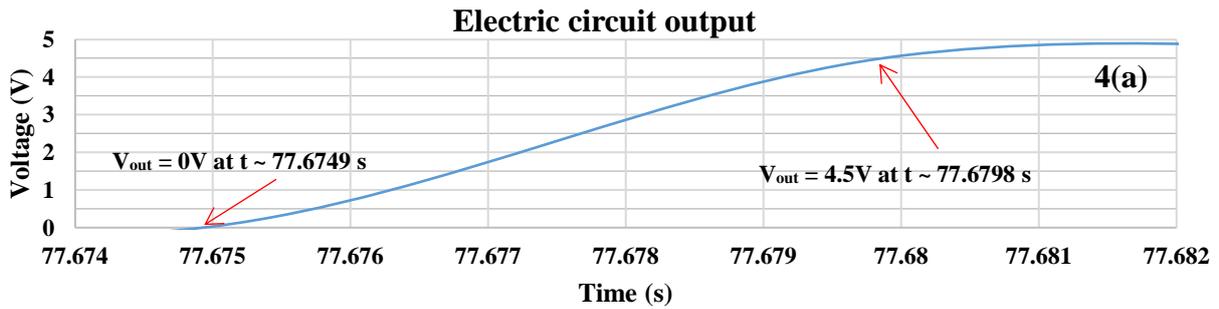
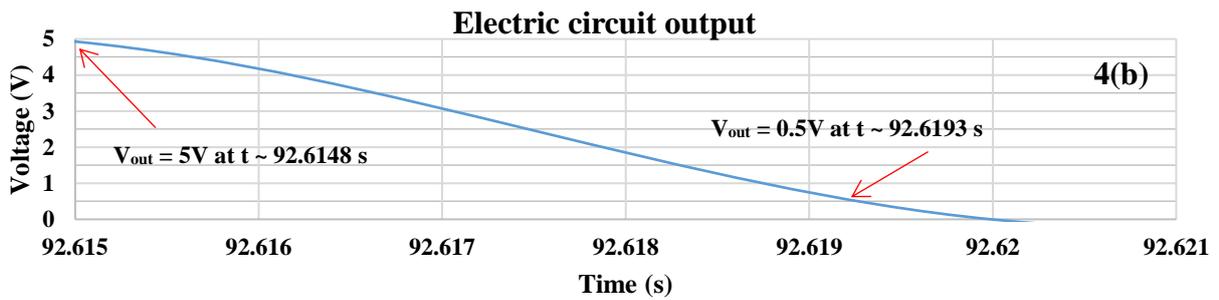

Fig. 4. Detailed analysis to estimate $T_{90}$ response-time during: (a) Immersion in PbLi; (b) Retrieval from PbLi

Further performance enhancements are possible with length reduction and use of needle-shaped tip to detect smaller bubbles with higher velocities. A relaxation in threshold limit (eg: $T_{50}$ or $T_{75}$ instead of $T_{90}$) will also allow detection of similar sized bubbles with higher velocities.

*C. Probe Performance at Low and Intermediate α*

Fabricated probe was exposed to PbLi-Ar two-phase vertical column with bulk PbLi temperature of 390°C and varying α. Obtained voltage pulses are normalized in binary, with logic HIGH (1) assigned to samples having magnitude ⩾ 80% of maximum output while logic LOW (0) assigned to remaining samples. Ratio of number of logic LOW samples to the total number of samples defines α for time-interval under consideration. Performance data for a continuous run of 20s (t = 0 to 20s), extracted from an experimental run of 200s duration, is presented in the form of simultaneous voltage and temperature trends (Fig.6(a)-(i)). Considering close agreement in redundant temperature trends from multivariable probe, trend for only one junction is shown and discussed as representative case. For ease of comparison, ordinate range is kept as 30°C and abscissa range as 2.5s with α mentioned at top-left corner. The first observation that Fig. 6(a)-(i) depict characteristics of bubbly flow occasionally tending towards slug flow [16, 21-22] without a clear distinction, is in close agreement to experimental observations from [9], where no stable slug bubbles could be identified in PbBi-$N_2$ flow owing to high shearing stress of LM flow induced by preceding bubbles. Fig. 6(i), extracted from same experimental run, depicts arrested gas-flow from the sparger with probe dipped in PbLi environment continuously. Fig. 7, derived from another experimental run, provides insights about PbLi-Ar flow at very low α (≠0) giving stark characteristics of well-dispersed Ar bubbles of varying diameters in PbLi continuum. Variation in experimentally measured two-phase bulk temperature using multivariable probe is observed to be inversely dependent on α, attributed to a coupled effect of forced convection and simultaneous decrease in the effective thermal conductivity of the two-phase mixture leading to a reduced heat-extraction from the tank walls. Obtained voltage data in agreement with temperature trends establishes reliable detection using fabricated probe. Restricted Ar flow (α =0 and 0.02), represented in Fig. 6(i) and Fig. 7, rapidly drives two-phase temperature towards bulk PbLi temperature. Although each bubble could be detected using electrical-conductivity principle, only a trend could be established through two-phase temperature measurements providing qualitative inference. Further adaptation of measurement probe is foreseen using micro-thermocouples with better response to quantify two-phase regime using temperature data alone.

*D. Bubble-Frequency and Average Bubble Residence-Time*

A single-tip probe could not provide information related to bubble velocity or impaction-chord in absence of a reference time frame. However, bubble-frequency ($f_b$), defined as the ratio of number of bubbles ($N$) impacting the probe to the given time duration ($T$), could be estimated [23] as follows:

$$f_b = N/T \quad (2)$$

$f_b$ can be further utilized to estimate average bubble residence-time ($T_b$), defined as average time a bubble takes to traverse through probe-tip [23], as follows:

$$T_b = \alpha / f_b \quad (3)$$

Considering cases from Fig. 6(a) and Fig. 6(d) with nearly same α, estimated $f_b$ are 4 bubble/s and 6 bubble/s, respectively. Such a difference could be explained by observation of unequal void distribution over 2.5s duration. Moreover, in a dispersed bubbly-flow regime, $f_b$ is expected to increase with an increase in α. However, high-surface tension of LMs leads to coalescence of bubbles resulting in gas-pockets with larger impaction-chords effectively reducing bubble count even with higher magnitude of α, as clearly observed for Fig. 6(h) with an estimated $f_b$ = 4.4 bubble/s. For the cases of Fig. 6(a) and Fig. 6(d) with similar α, estimated $T_b$ are ~118.75 ms and 79.5 ms, respectively, which also corroborates that even within similar α regimes, time-distribution of interaction of liquid and gas phases with the probe-tip varies significantly. For higher α case in Fig. 6(h), $T_b$ = 140.45 ms depicts the effect of bubble coalescence in accordance to the arguments made above. At very low α (Fig.7) regimes, estimated parameters of 0.6 bubble/s and 38.33 ms highlight a dispersed bubbly flow regime.

*E. Probe Performance at High α (In-box LOCA)*

In-box LOCA is one of the postulated accidental events where high-pressure secondary coolant (He/$H_2O$) enters PbLi loop pressurizing the complete PbLi system. A recent simulation study for PbLi-He two-phase flow [16] has considered α in the range of 0.1-0.9 with mixture velocities from 0.0001 m/s to 100 m/s in rectangular duct geometries of varying widths. In the present experimental study, controlled Ar flow varying from 1.5-3.5 slpm, corresponding to superficial gas velocities ($V_g$) from 0.0115 m/s to 0.0269 m/s (at standard conditions), has been injected using a DMFC to achieve α ~ 0.95 over continuous durations as long as 200s within bulk PbLi temperature of 400°C. Fig. 8 represents a sample of 10s corresponding to $V_g$ = 0.0192 m/s. Flow-pattern inferred from pulsed voltage output clearly establishes localized two-phase regime tending towards an annular flow with $f_b$ = 1.3 bubble/s and $T_b$ = 726.92 ms. In contrast to bubbly flow, high $T_b$ is indicative of dispersed PbLi phase within Ar gas continuum near the probe-tip. This observation is also in agreement with the annular flows reported under [16] at relevant α. The average temperature measured by multivariable probe and PbLi bulk temperature sensor over the presented time duration matches within ±1°C to corresponding averages over complete test-duration of 250s, justifying the analysed sample being representative case. Continuous exposure duration of the fabricated probe in high α (> 0.9) PbLi-Ar environment was over 2h. No signal distortions related to coating degradation were observed [24] and insulation integrity was corroborated with $I_L$ < 1nA at 275 VDC post-exposure. During the preliminary investigations, probe was exposed to temperature gradients ~ 300°C within a few seconds (during immersion) and pressure gradients upto 1 bar(g) during sparger tests. No adverse effects of such gradients were observed. Image of the probe after retrieval is shown in Fig. 9. The silvery lustrous PbLi deposited on immersed portion is due to installation constraint, requiring complete retraction of probe before retrieval of sparger.

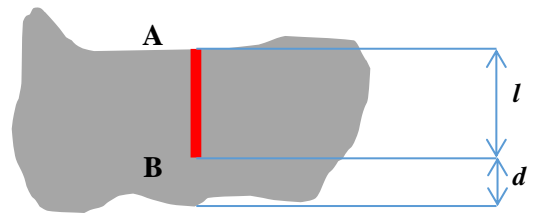

Fig. 5. Electrical-conductivity based detection criteria towards detection of an irregular bubble

Probe was chemically cleaned and re-tested for satisfactory functioning in PbLi-Ar columns.

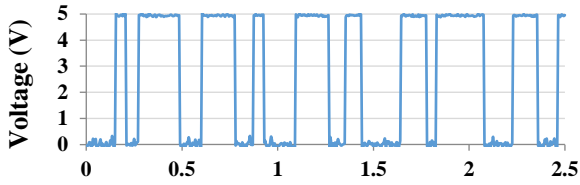
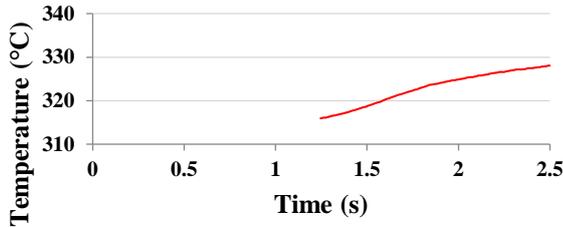

6(a)

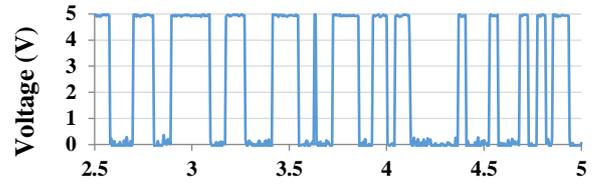
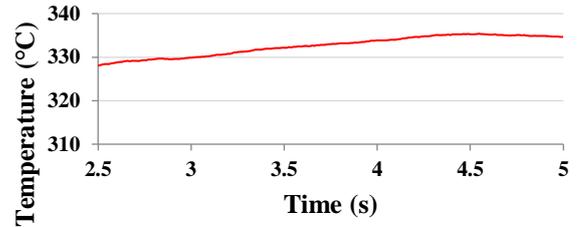

6(b)

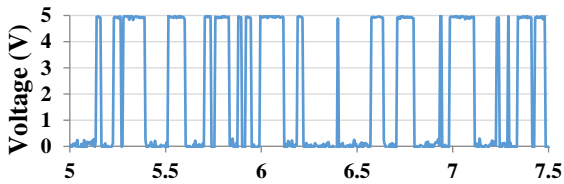
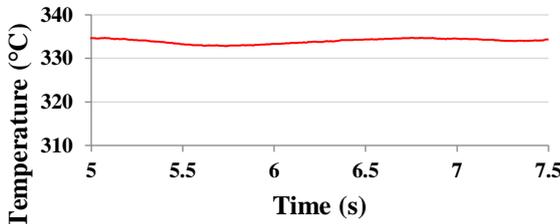

6(c)

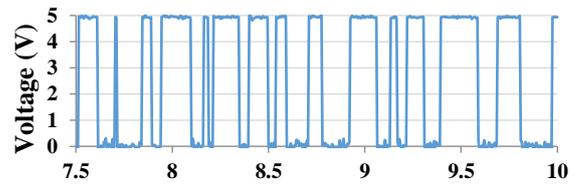
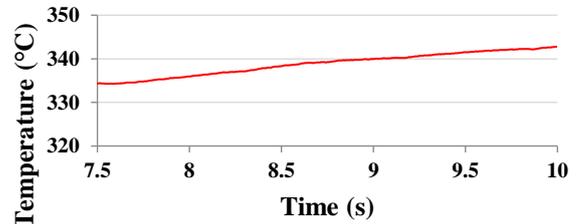

6(d)

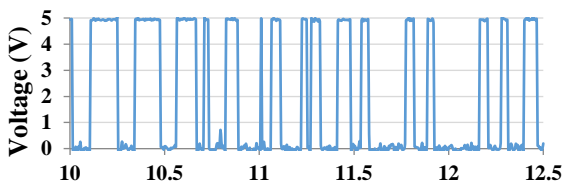
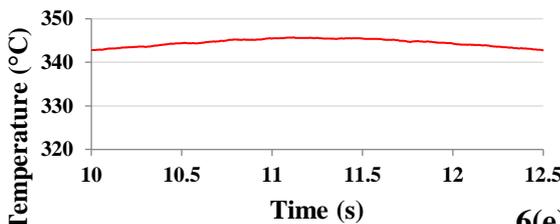

6(e)

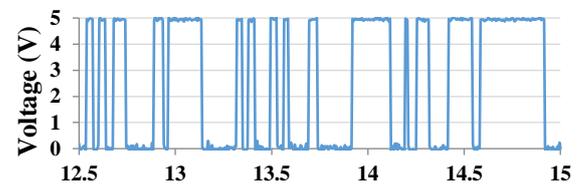
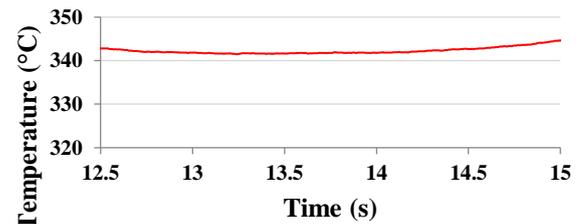

6(f)

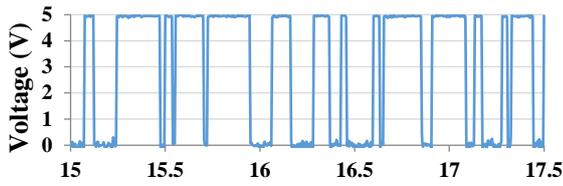
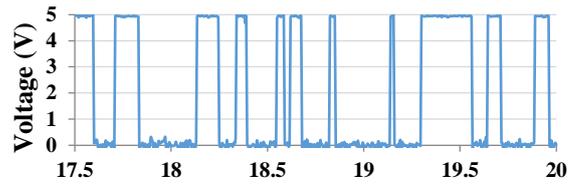
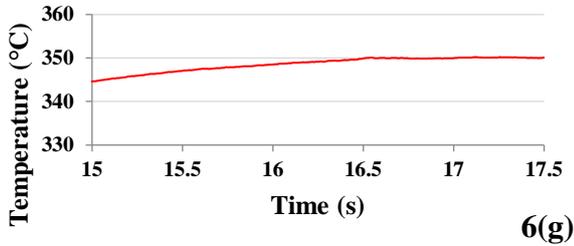
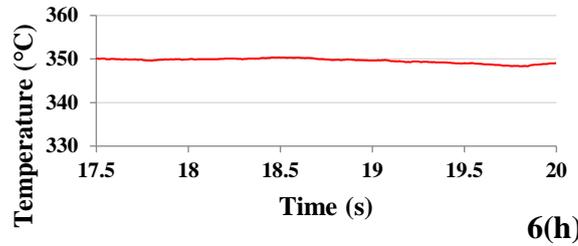

6(g)　　6(h)

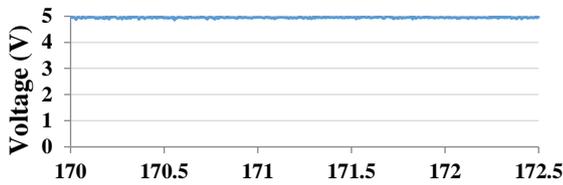
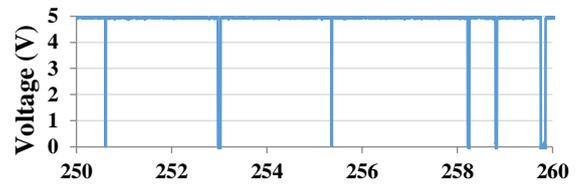
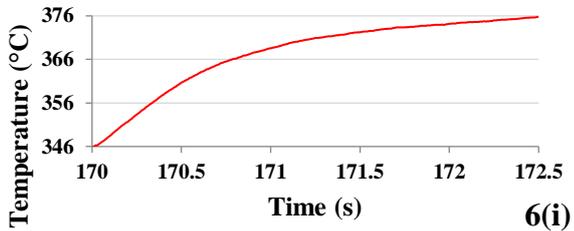
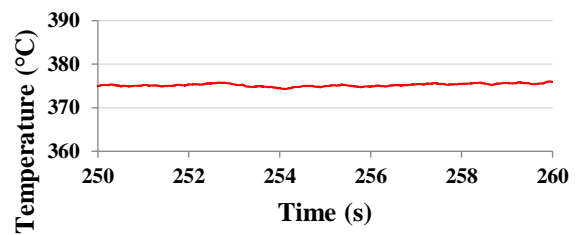

6(i)

Fig. 6(a)-(i). Voltage and temperature data for varying α in PbLi-Ar two phase column

Fig. 7. Performance of probe at very low (≠0) α

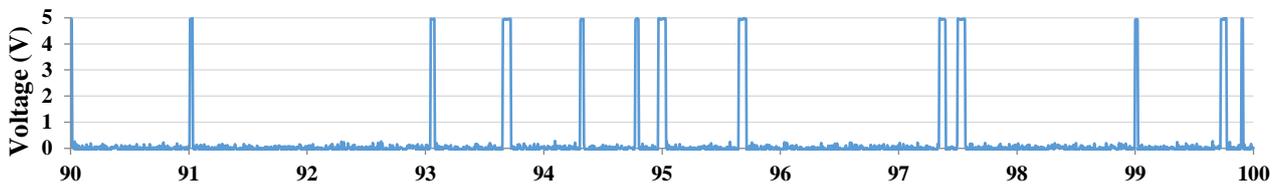
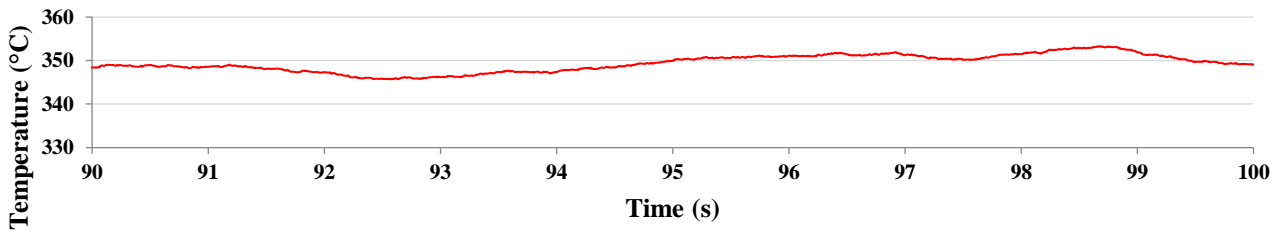

Fig. 8. Multivariable probe response for Argon $V_g$ = 0.0192 m/s

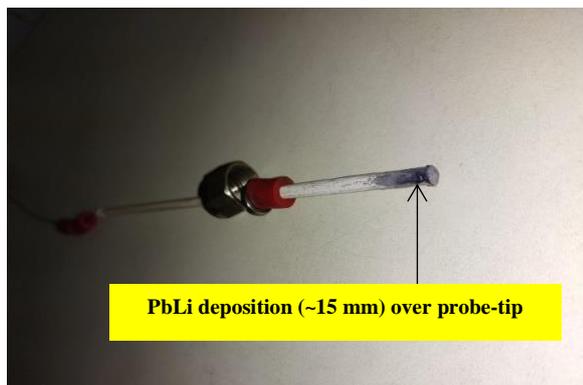

Fig. 9. Multivariable probe post-exposure to PbLi-Ar two-phase environment

IV. CONCLUSIONS

A multivariable probe based on electrical-conductivity principle employing simultaneous bulk temperature measurements was fabricated, using high-purity electrically insulating $Al_2O_3$ coating, to experimentally study PbLi-Ar two-phase flow with bulk temperature upto 400°C. Developed probe was calibrated and validated under different flow regimes like single-phase, bubbly flow, transition and localized annular flows with void-fractions upto 0.95. Conclusive bubble detection criterion was discussed to correlate bubble-velocity and impaction-chord length for a given threshold. Electrical-conductivity based scheme provides reliable and quantitative detection while coherent temperature trends provide qualitative insights relevant to existing flow regime. Bubble-frequency and average bubble residence-time were calculated from the measured data. Experimental observations suggest increase in gas-phase fraction ($\alpha > 0.6$) may result in coalescence of smaller bubble pockets leading to an increased bubble residence-time. In contrast to water-air flows, present study corroborates absence of stable slug bubbles, ascribed to shearing stresses generated by induced liquid-metal flow. A clear demarcation of flow regime requires fully developed flow, which was limited in the present study. To minimize effects related to probe-bubble interactions (like flow-geometry modification, bubble dynamics, etc.), miniaturization is envisaged using fine-diameter thermocouples. Although time-averaged α is relevant to study localized phenomena, ability to measure area-averaged void-fractions is of extreme importance. In this view, a sensor-array is planned for implementation in PbLi-Ar/He vertical columns. Developed probe is expected to be an indispensable tool enabling experimental studies at high-temperature providing substantiation for software tools under development towards predicting postulated events in a TBM and/or fusion machine employing advanced liquid-metal based breeding-blanket concepts.